\input{aipcheck}

\documentclass[
    ,final            
  ]
  {aipproc}

\layoutstyle{8x11double}

\usepackage{amssymb}
\usepackage{amsmath}
\usepackage{axodraw}
\usepackage{graphpap}

\newcommand{\tw}[1]{\widehat{T}_{#1}}
\newcommand{\mb}{\overline{m}}
\newcommand{\zsi}{$Z_6$--I}

\begin{document}

\title{Fermion masses and mixings from heterotic orbifold models%
  \footnote{Talk presented 
    at PASCOS-05, Gyeongju,
    May 30--Jun 4, 2005; to appear in the proceedings.}
}


%

\author{Jae-hyeon~Park}{
  address={School of Physics, KIAS, Cheongnyangni-dong, Seoul 130-722, Korea}
}

\begin{abstract}
  We search for a possibility of getting realistic fermion mass ratios
  and mixing angles from
  renormalizable couplings on the \zsi\ heterotic orbifold
  with one pair of Higgs doublets.
  In the quark sector, we find cases with reasonable
  $\mb_c/\mb_t$, $\mb_s/\mb_b$, and $V_{cb}$,
  if we ignore the first family.
  In the lepton sector, we can fit the charged lepton mass ratios,
  the neutrino mass squared difference ratio, and the lepton mixing angles,
  considering all three families.
\end{abstract}

\maketitle




\begin{picture}(0,0)
  \put(400,190){KIAS--P05053}
\end{picture}%
In heterotic string theory, there are 26 bosonic left-moving
degrees of freedom and 10 supersymmetric right-moving degrees of freedom.
Among these, four left-movers and four right-movers are the
observed spacetime.
The rest are compactified.
Among the left-movers, 16 of them are responsible for the
internal gauge symmetry.
Remaining six left-movers and six right-movers can serve as
an origin of the flavor structure of the Yukawa couplings.
Orbifold is commonly used for the geometry of these internal six dimensions.
On an orbifold, there are fixed points, and
a twisted closed string ground state is attached to
each of these fixed points.
A trilinear string scattering amplitude of three of these states
is written as
\begin{equation*}
  \int \mathcal{D}X\ e^{-S} \sigma_1 (z_1) \sigma_2 (z_2) \sigma_3 (z_3)
  = Z_\mathrm{qu} \sum_{\langle X_\mathrm{cl}\rangle}
  e^{- S_\mathrm{cl}} ,
\end{equation*}
where $\sigma_i$ represents a twist field creating the
appropriate twisted ground state.
The quantum part $Z_\mathrm{qu}$ in the right-hand side
is a global factor for all the couplings with a given
twist structure, and the flavor structure essentially comes from
the classical part.
The sum is over all the classical string configurations, and the
classical action $S_\mathrm{cl}$ is given by the world sheet area.
Therefore this amplitude depends on the distances
among the three fixed points which are determined by
the relative positions of the fixed points and
the volumes of the tori comprising the internal dimensions.
Since the amplitude has an exponential suppression factor
depending on the volumes,
one can hope to use this to account for the hierarchical
structure of fermion masses and mixings.

\begin{figure}[tbp]
  \centering
  \includegraphics[width=1.7in]{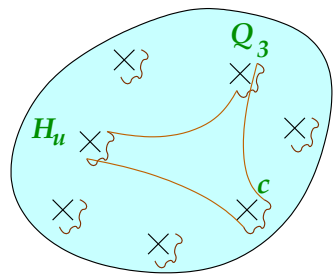}
  \caption{As an example, the up-type Higgs doublet $H_u$,
   the third generation $SU(2)$ doublet quark $Q_3$,
   and the $SU(2)$ singlet charm quark are assigned to three different
  twisted close string ground states.
  The Yukawa matrix element $(Y_u)_{32}$ given by
  their trilinear string scattering amplitude, is proportional
  to the area of the world sheet shown in the drawing.}
  \label{fig:pict}
\end{figure}

The overall picture of this work is illustrated in
Fig.~\ref{fig:pict}.
We associate each of the matter fields such as
quarks, leptons, and Higgses, to one or more of the fixed points.
This association is done by hand, which means that
we do not specify how a Standard-like model is obtained
by fixing gauge shifts and Wilson lines \cite{Ibanez:1987sn}.
Therefore, this work is a sort of model-independent analysis
performed on a specific type of orbifold, \zsi\ in the present case.
Once the field assignment is fixed, we can compute the
Yukawa couplings of the quarks or leptons as functions
of the moduli $R^2_{1,2,3}$,
describing the sizes of the three two-dimensional tori.
Then, we fit quark or lepton mass ratios and mixing angles
varying $R^2_{1,2,3}$.
We repeat this procedure for each possible assignment,
looking for a case which can reproduce the observed
mass ratios and mixings for well-chosen values of the moduli.


All the six-dimensional orbifolds that give rise to
$N=1$ four-dimensional supersymmetry have been classified.
Among these, prime orbifolds such as $Z_3$ and $Z_7$
are not useful for our purpose.
Suppose that three states attached to the three fixed points $f_{1,2,3}$
form a Yukawa coupling.
Given $f_1$ and $f_2$,
the space group selection rule on a prime orbifold
uniquely determines $f_3$ which can couple to $f_1$ and $f_2$.
Because of this property,
a Yukawa matrix from renormalizable couplings
either is diagonal, or has a zero eigenvalue.
If we avoid a massless quark, we are lead to have a trivial CKM
matrix equal to identity.
Therefore, we do not consider a prime orbifold.
On a non-prime orbifold, $f_3$ is not uniquely determined,
and nontrivial mixing is possible.
We consider the 
\zsi\ orbifold because the other orbifolds
have smaller number of available twisted states
or smaller number of parameters which can be tuned
to adjust fermion mass ratios and mixing angles.
However, this does not necessarily mean that 
all the other non-prime orbifolds are not
phenomenologically viable.
Also, one should keep in mind that
nonrenormalizable couplings may always contribute to Yukawa couplings.
This is the reason why
string phenomenology on prime orbifolds
is not useless.

Before describing the \zsi\ orbifold,
let us first look at two-dimensional orbifolds which will be
used to construct it.
A 2D $Z_3$ orbifold looks like Fig.~\ref{fig:2dz3z6}~(a).
It is a torus modded by $120^\circ$ of rotation.
It has the following fixed points and
their respective twisted ground states:
\begin{equation}
  \begin{aligned}
  g^{(0)}_{Z_3,1}&=(0,0)     &&\rightarrow && | g^{(0)}_{Z_3,1} \rangle , \\
  g^{(1)}_{Z_3,1}&=(1/3,2/3) &&\rightarrow && | g^{(1)}_{Z_3,1} \rangle , \\
  g^{(2)}_{Z_3,1}&=(2/3,1/3) &&\rightarrow && | g^{(2)}_{Z_3,1} \rangle .
  \end{aligned}
\end{equation}
The parenthesized coordinates are in the units of the basis vectors
$e_1$ and $e_2$.
Another relevant 2D orbifold is the $Z_6$ orbifold,
shown in Fig.~\ref{fig:2dz3z6}~(b).
It is a torus modded by $60^\circ$ of rotation.
Let $\theta$ denote this rotation.
The structure of fixed points and the twisted states
on this orbifold is more involved
since it has $\theta^2$-twisted and $\theta^3$-twisted sectors
as well as $\theta$-twisted sector.
They are summarized in Table~\ref{tab:2dz6fps}.
A physical state should be a $\theta$ eigenstate.
Taking linear combinations of the states attached to
the fixed points \cite{kobayashi}, 
one can get the $\theta$ eigenstates,
\begin{equation}
  \label{eq:2dz6states}
  \begin{aligned}
      &| g^{(0)}_{Z_6,1} \rangle,  \\
      &| g^{(0)}_{Z_6,2} \rangle, \
      | g^{(1)}_{Z_6,2}; \pm 1 \rangle \equiv | g^{(1)}_{Z_6,2} \rangle 
        \pm | g^{(2)}_{Z_6,2} \rangle , \\
      &| g^{(0)}_{Z_6,3} \rangle, \
      | g^{(1)}_{Z_6,3}; \gamma \rangle \equiv  | g^{(1)}_{Z_6,3} \rangle 
      + \gamma  | g^{(2)}_{Z_6,3} \rangle + \gamma^2  | g^{(3)}_{Z_6,3} \rangle, 
  \end{aligned}
\end{equation}
where $\gamma = 1, \omega, \omega^2$ with $\omega = e^{2 \pi i / 3}$.
The 6D \zsi\ orbifold is a direct product of two 2D
$Z_6$ orbifolds and one 2D $Z_3$ orbifold.
The fixed points are given as direct products of those on the 2D orbifolds,
and therefore the corresponding twisted states follow in the same manner.
Due to the point group selection rule and $H$-momentum conservation,
one has two types of possible Yukawa couplings,
\begin{equation*}
    {\tw{1}\tw{2}\tw{3}} , \quad
    {\tw{2}\tw{2}\tw{2}} ,
\end{equation*}
where $\tw{1}$, $\tw{2}$, and $\tw{3}$ are
states from the $\theta$-, $\theta^2$-, and $\theta^3$-twisted sectors,
respectively.
Each of these states is
a direct product of two states from~(\ref{eq:2dz6states})
with the corresponding twist, and
their concrete expressions can be found in~\cite{Ko:2004ic,Ko:2005sh}.
One can show that the 2D $Z_3$ orbifold contributes only
to the overall factor of a Yukawa matrix,
thus being irrelevant to fermion mass
ratios and mixings.
However, this part
can be used to scale a Yukawa matrix to a desirable order
of magnitude.
For example, $\tan \beta$ can be changed by scaling either
$Y_u$ or $Y_d$.
\def\axoscale{0.4 }
\setlength{\unitlength}{.4pt}
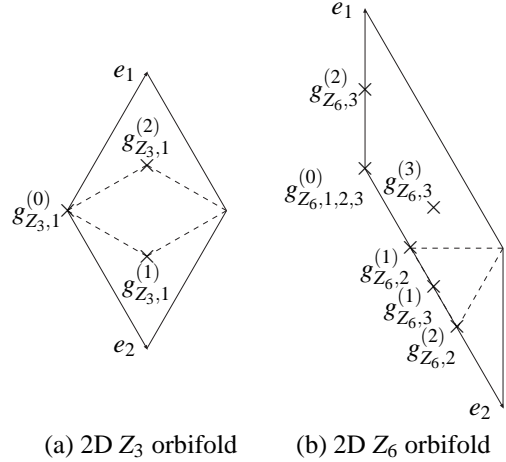
\begin{figure}[tbp]
  \parbox{3.125in}{%
  \centering
\begin{picture}(200,400)(-100,-200)
    \LongArrow(-75,0)(0,130)
    \LongArrow(-75,0)(0,-130)
    \Text(-10,130)[r]{$e_1$}
    \Text(-10,-130)[r]{$e_2$}
    \Line(0,130)(75,0)
    \Line(0,-130)(75,0)
    \DashLine(-75,0)(0,43){5}
    \DashLine(-75,0)(0,-43){5}
    \DashLine(75,0)(0,43){5}
    \DashLine(75,0)(0,-43){5}
    \Text(-75,0)[c]{$\times$}
    \Text(-80,0)[r]{$g^{(0)}_{Z_3,1}$}
    \Text(0,43)[c]{$\times$}
    \Text(0,50)[b]{$g^{(2)}_{Z_3,1}$}
    \Text(0,-43)[c]{$\times$}
    \Text(0,-50)[t]{$g^{(1)}_{Z_3,1}$}
  \end{picture}
\qquad
\begin{picture}(200,400)(-50,-240)
    \LongArrow(0,0)(0,150)
    \LongArrow(0,0)(130,-225)
    \Line(0,150)(130,-75)
    \Line(130,-225)(130,-75)
    \Text(-10,150)[r]{$e_1$}
    \Text(120,-230)[r]{$e_2$}
    \DashLine(130,-75)(43,-75){5}
    \DashLine(130,-75)(87,-150){5}
    \DashLine(43,-75)(87,-150){5}
    \Text(0,0)[c]{$\times$}
    \Text(0,0)[rt]{${g^{(0)}_{Z_6,1,2,3}}$}
    \Text(43,-75)[c]{$\times$}
    \Text(43,-75)[rt]{$g^{(1)}_{Z_6,2}$}
    \Text(87,-150)[c]{$\times$}
    \Text(87,-150)[rt]{$g^{(2)}_{Z_6,2}$}
    \Text(65,-112)[c]{$\times$}
    \Text(65,-112)[rt]{$g^{(1)}_{Z_6,3}$}
    \Text(0,75)[c]{$\times$}
    \Text(-5,75)[r]{$g^{(2)}_{Z_6,3}$}
    \Text(65,-37)[c]{$\times$}
    \Text(65,-31)[rb]{$g^{(3)}_{Z_6,3}$}
  \end{picture}
  \\
  (a) 2D $Z_3$ orbifold \qquad (b) 2D $Z_6$ orbifold
  \hspace{2ex}
  }
  \caption{Two-dimensional $Z_3$ and $Z_6$ orbifolds.
  A cross marks a fixed point.
  The region inside dashed lines is the fundamental domain of
  the orbifold.}
  \label{fig:2dz3z6}
\end{figure}
\begin{table}[bp]
  \centering
  \begin{tabular}{ll}
    \hline
    Under & Fixed points are \\ \hline 
    $
    \theta \text{($= 60^\circ$ rot.)}
    $
    & $g^{(0)}_{Z_6,1} = (0,0)$ \\ \hline 
    $\theta^2$
    & 
    $
    \begin{aligned}
      g^{(0)}_{Z_6,2} &= (0,0), \quad g^{(1)}_{Z_6,2} = (0,1/3), \\
      g^{(2)}_{Z_6,2} &= (0,2/3) 
    \end{aligned} $
    \\ \hline 
    $\theta^3$ &
    $
    \begin{aligned}
      g^{(0)}_{Z_6,3} &= (0,0),  & g^{(1)}_{Z_6,3} &= (0,1/2), \\
      g^{(2)}_{Z_6,3} &= (1/2,0), & g^{(3)}_{Z_6,3} &= (1/2,1/2) 
    \end{aligned} $
    \\ \hline
  \end{tabular}
  \caption{Two-dimensional $Z_6$ orbifold has three different
    twisted sectors.
  The fixed points which are invariant under each twist are shown here.
  The coordinates in the parentheses are with respect to 
  $e_1$ and $e_2$ in Fig.~\ref{fig:2dz3z6}~(b).}
  \label{tab:2dz6fps}
\end{table}

In this work, we assume that there exists a model
based on the \zsi\ heterotic orbifold realizing
the following points:
\begin{itemize}
\item $SU(3) \times SU(2) \times U(1)_Y$ observable gauge group.
\item Three families of quarks and leptons.
\item One family of Higgs doublets, $H_u$ and $H_d$.
\item All matter fields come from twisted sectors.
\end{itemize}
Among these, the last point is due to the fact that
it is very hard to get
hierarchical fermion masses using untwisted sector fields.

\addtocounter{table}{1}%
\newcommand{\rowq}[9]{$\tw{#1}^{(#2)}$&$\tw{#1}^{(#3)}$&$\tw{#4}^{(#5)}$&$\tw{#4}^{(#6)}$&$\tw{#7}^{(#8)}$&$\tw{#7}^{(#9)}$}%
\begin{table}[tb]
  \centering
\begin{tabular}{cl@{}l@{}l@{}l@{}l@{}l@{}l@{}lccccc}
      \hline
      Class & $Q_2$&$Q_3$&$u^c_2$&$u^c_3$&$d^c_2$&$d^c_3$&$H_u$&$H_d$&$R_1^2$&$R_2^2$&${\mb_c}/{\mb_t}$&${\mb_s}/{\mb_b}$&$V_{cb}$ \\
      \hline
      1 & \rowq{2}{2}{4}{3}{3}{2}{3}{1}{3}&$\tw{1}      $&$\tw{1}      $ & 27.8 & 107  & 0.0038 & 0.029 & 0.041 \\
      2 & \rowq{3}{2}{4}{2}{3}{2}{2}{1}{3}&$\tw{1}      $&$\tw{1}      $ & 24.0 & 150  & 0.0038 & 0.032 & 0.041 \\
      3 & \rowq{2}{1}{4}{3}{2}{4}{2}{2}{4}&$\tw{1}      $&$\tw{2}^{(4)}$ & 196  & 316  & 0.0038 & 0.019 & 0.042 \\
      4 & \rowq{2}{2}{4}{2}{2}{3}{3}{1}{4}&$\tw{2}^{(4)}$&$\tw{1}      $ & 416  & 226  & 0.0040 & 0.035 & 0.035 \\
      5 & \rowq{2}{2}{4}{2}{2}{4}{2}{3}{2}&$\tw{2}^{(4)}$&$\tw{2}^{(4)}$ & 368  & 400  & 0.0038 & 0.029 & 0.041 \\
      \hline
    \multicolumn{11}{c}{Central values from measurements \cite{quarkdata}} & 0.0038 & 0.025 & 0.041 \\
    \hline
  \end{tabular}%
  \caption{An example field assignment from each class.
    Values of the moduli $R_1^2$ and $R_2^2$ which lead to
    the best fit of the quark mass ratios and $V_{cb}$ are also shown.
    Central mass ratio values in the last row are
    from the running quark masses at $m_W$ scale.
    Meaning of each symbol denoting a state can be found
    in~\cite{Ko:2004ic}.}
  \label{tab:quarkstates}
\end{table}%
\addtocounter{table}{-2}%
\begin{table}[tbp]
  \centering
  \begin{tabular}{cccccc}
    \hline
    Class & $Q$ or $L$ & $u^c$ or $N$ & $d^c$ or $e^c$ & $H_u$ & $H_d$ \\
    \hline
    1 & $\tw{2}$ & $\tw{3}$ &  $\tw{3}$ & $\tw{1}$ & $\tw{1}$ \\
    2 & $\tw{3}$ & $\tw{2}$ &  $\tw{2}$ & $\tw{1}$ & $\tw{1}$ \\
    3 & $\tw{2}$ & $\tw{3}$ &  $\tw{2}$ & $\tw{1}$ & $\tw{2}$ \\
    4 & $\tw{2}$ & $\tw{2}$ &  $\tw{3}$ & $\tw{2}$ & $\tw{1}$ \\
    5 & $\tw{2}$ & $\tw{2}$ &  $\tw{2}$ & $\tw{2}$ & $\tw{2}$ \\
    \hline
  \end{tabular}
  \caption{Five classes of assignments.}
  \label{tab:class}
\end{table}%
\addtocounter{table}{1}%
Let us first discuss the quark sector \cite{Ko:2004ic}.
We assign each of $Q_{1,2,3}$, $u^c_{1,2,3}$, $d^c_{1,2,3}$, $H_u$,
and $H_d$ to one of the twisted states. 
In order to get nontrivial quark mass ratios and mixings, we are lead
to consider one of the five classes of assignments shown in
Table~\ref{tab:class}.
In each of these classes, we examine every possibility of
field assignment, for which
we perform fitting of
$\mb_u/\mb_t$, $\mb_c/\mb_t$,
$\mb_d/\mb_b$, $\mb_s/\mb_b$,
$V_{us}$, $V_{cb}$, and $V_{ub}$,
varying $R_1^2$ and $R_2^2$.
Recall that $R_3^2$ is irrelevant to mass ratios and mixings.
We ignore quark mass running between the string scale and the weak scale.
If we try to fit all of the mass ratios and the CKM matrix elements above,
we do not find a satisfactory result.
Ignoring the first family quarks, however,
we can get values of $\mb_c/\mb_t$, $\mb_s/\mb_b$,
and $V_{cb}$, which are fairly close to the central values from measurements.
We quote one instance of fit from each assignment class
in Table~\ref{tab:quarkstates}.
There are a number of other instances in addition to the one
shown in the table for each class.
To our knowledge, this is the first
work that showed a possibility
of getting
realistic mixing angles
from renormalizable couplings
in string models with one pair of Higgs fields.


Now we turn to the lepton sector \cite{Ko:2005sh}.
The analysis procedure here almost parallels that
for the quark sector.
Computation of the lepton Yukawa couplings is performed
in the same way except that we should replace $(Q, u^c, d^c)$
by $(L, N, e^c)$.
A complication is that the neutrino masses
may be of Majorana type, in addition to Dirac type which
is a direct analogy of quark masses.
One customarily incorporates the seesaw mechanism for Majorana
neutrino masses to explain lightness of neutrinos.
In this work, we consider two neutrino mass generation mechanisms:
Dirac mass scenario, and seesaw scenario with the right-handed
neutrino mass matrix proportional to a unit matrix.
For each of these scenarios, we assign lepton and Higgs fields
to the twisted states and fit
    $m_e/m_\tau$, $m_\mu/m_\tau$,
    $\Delta m^2_{31} / \Delta m^2_{21}$,
    $\sin^2 \theta_{12}$, $\sin^2 \theta_{23}$, and $\sin^2 \theta_{13}$.

The result in the Dirac scenario is summarized in
Table~\ref{tab:diracstates}.
\begin{table}[tb]
  \centering
  \begin{tabular}{ccccc}
    \hline
    Class & 
    $\Delta m^2_{31}/\Delta m^2_{21}$ & 
    $\sin^2 \theta_{12}$ & $\sin^2 \theta_{23}$ & $\sin^2 \theta_{13} $ \\
    \hline
    1 & 
    $\sim 100$ & $\lesssim 10^{-5}$ & $\lesssim 10^{-2}$ & $\lesssim 10^{-7}$\\
    2 & 
    $\sim 100$ & $\lesssim 10^{-5}$ & $\lesssim 10^{-2}$ & $\lesssim 10^{-7}$\\
    3 & 
    $\gtrsim 1.4$ &  &  &  \\
    4 & 
    14 & 0.38 & 0.70 & $6.3 \times 10^{-6}$ \\
    5 & 
    $\sim 28$ & $\le 0.09$ & & $\lesssim 10^{-2}$ \\
    \hline
    Central values \cite{Maltoni:2004ei}
    & 
    27 & 0.30 & 0.50 & 0.000 \\
    \hline
  \end{tabular}%
  \caption{Characteristics of each class in the Dirac
    neutrino case.
    Typical behavior of each quantity is described
    for combinations resulting in relatively good fits
    in a given class except Class 4.
    The row corresponding to Class 4 shows the best fit.
    We omit $m_e/m_\tau$ and $m_\mu/m_\tau$ because
    they can be fit in all the classes.}
  \label{tab:diracstates}
\end{table}
\begin{table}[tb]
  \centering
  \begin{tabular}{ccccc}
    \hline
    Class & 
    $\Delta m^2_{31}/\Delta m^2_{21}$ & 
    $\sin^2 \theta_{12}$ & $\sin^2 \theta_{23}$ & $\sin^2 \theta_{13} $ \\
    \hline
    1 & 
    $\sim 6000$& $\lesssim 10^{-5}$ & $\lesssim 10^{-2}$ & $\lesssim 10^{-7}$\\
    2 & 
     $\sim 7000$&$\lesssim 10^{-5}$ & $\lesssim 10^{-2}$ & $\lesssim 10^{-7}$\\
    3 & 
    $\gtrsim 2$ &  &  &  \\
    4 & 
    29 & 0.32 & 0.48 & $3.6 \times 10^{-6}$ \\
    5 & 
    $\sim 28$ & $\le 0.09$ & & $\lesssim 10^{-2}$ \\
    \hline
    Central values \cite{Maltoni:2004ei}
    &
    27 & 0.30 & 0.50 & 0.000 \\
    \hline
  \end{tabular}%
  \caption{Characteristics of each class in the seesaw case.
    Typical behavior of each quantity is described
    for combinations resulting in relatively good fits
    in a given class.
    The row corresponding to Class 4 shows the best fit.
    We omit $m_e/m_\tau$ and $m_\mu/m_\tau$ because
    they can be fit in all the classes.}
  \label{tab:seesawstates}
\end{table}
In Classes 1, 2, 3, and 5, we do not find a good fit of the neutrino
mass squared difference ratio and mixing angles.
Approximate (in)equalities in these classes show the typical behavior
of each quantity for assignments with relatively low $\chi^2$.
In contrast to the other classes,
Class 4 leads to promising results.
It is notable that we can fit all the above six observables
tuning only two parameters $R_1^2$ and $R_2^2$, in this class.
One example assignment is as follows:
\begin{equation*}
  \begin{aligned}
(L_1,L_2,L_3) &= (\tw{2}^{(2)},\tw{2}^{(3)},\tw{2}^{(4,-1)}), \\
(N_1,N_2,N_3) &= (\tw{2}^{(2)},\tw{2}^{(4,1)},\tw{2}^{(4,-1)}), \\
(e^c_1,e^c_2,e^c_3) &= (\tw{3}^{(1)},\tw{3}^{(2)},\tw{3}^{(4,1)}), \\
(H_u,H_d) &= (\tw{2}^{(2)},\tw{1}).
  \end{aligned}
\end{equation*}
Meaning of each symbol on the right-hand sides is
available in~\cite{Ko:2005sh}.
This assignment results in the fit shown in Table~\ref{tab:diracstates}
for $(R^2_1,R^2_2) = (26.0, 20.6)$.
In this scenario, smallness of neutrino masses should be accounted for
by small Yukawa couplings.
For this, one can use the 2D $Z_3$ orbifold.
For example, $L$, $N$, and $H_u$ can be put at three different
fixed points on the $Z_3$ orbifold,
with three families of $L$ or $N$ gathered at a single point.
If the size of this orbifold is taken to be big enough so that
$R_3^2 \sim 1000$, one can have a sufficient suppression factor for
the neutrino Yukawa couplings.

In the seesaw scenario, we assume that
the right-handed neutrino mass matrix is
proportional to an identity matrix.
Therefore, the lepton mixing angles are essentially determined by
the Yukawa couplings as in the Dirac scenario.
One difference is that a physical neutrino mass eigenvalue
is proportional to the square of a Yukawa matrix eigenvalue.
This enhances the mass squared difference ratio relative to
that in the Dirac scenario.
Indeed, $\Delta m^2_{31}/\Delta m^2_{21}$ in
Table~\ref{tab:seesawstates} for an example assignment in Class 4
is larger and hence is closer to the central value
than in Table~\ref{tab:diracstates}.
This fit was obtained using the following assignment:
\begin{equation*}
\begin{aligned}
(L_1,L_2,L_3) &= (\tw{2}^{(1)},\tw{2}^{(2)},\tw{2}^{(4,1)}), \\
(N_1,N_2,N_3) &= (\tw{2}^{(2)},\tw{2}^{(3)},\tw{2}^{(4,1)}), \\
(e^c_1,e^c_2,e^c_3) &= (\tw{3}^{(1)},\tw{3}^{(2)},\tw{3}^{(4,1)}), \\
(H_u,H_d) &= (\tw{2}^{(2)},\tw{1}),
\end{aligned}  
\end{equation*}
for $(R_1^2, R_2^2) = (22.5, 26.0)$.
In this case, we need the scale of the right-handed neutrino mass
$M_N \sim 10^{15}$ GeV
for the neutrino masses to be of the right order of magnitude.
As in the Dirac scenario, the other classes do not lead to
an acceptable fit of the observables.
Let us remark that the predicted value of $\theta_{13}$ is
vanishingly small and the neutrino mass spectrum shows
normal hierarchy both in the seesaw scenario
and in the Dirac scenario.


In conclusion, we systematically searched for possibilities
to get realistic fermion
mass ratios and mixings from the \zsi\ heterotic orbifold.
We assumed that Yukawa matrices of quarks and leptons
arise from renormalizable couplings
with one family of Higgses.
In the quark sector,
we could obtain reasonable values of $\mb_c/\mb_t$, $\mb_s/\mb_b$, 
\emph{and} $V_{cb}$ ignoring the first family, although
we failed to get an acceptable fit in the
three family analysis.
In the lepton sector,
we could fit the six observables of
$m_e/m_\tau$, $m_\mu/m_\tau$,
$\Delta m^2_{31}/\Delta m^2_{21}$,
$\theta_{12}$, $\theta_{23}$, and $\theta_{13}$,
by adjusting only two moduli parameters
in either the Dirac or the seesaw scenario.


\vspace{3ex}
  The author is grateful to Pyungwon~Ko and Tatsuo~Kobayashi for the
  pleasurable collaborations.


\end{document}